\documentclass{article}

\usepackage[preprint]{neurips_2025}

\usepackage[utf8]{inputenc} 
\usepackage[T1]{fontenc}    
\usepackage{hyperref}       
\usepackage{url}            
\usepackage{booktabs}       
\usepackage{amsfonts}       
\usepackage{nicefrac}       
\usepackage{microtype}      
\usepackage{xcolor}         

\usepackage{amsmath}
\usepackage{array}
\usepackage{graphicx}

\title{PestMA: LLM-based Multi-Agent System for Informed Pest Management}

\author{%
  Hongrui Shi \\
  School of Computer Science\\
  University of Sheffield\\
  Sheffield S10 2TN, UK \\
  \texttt{alex.yalishanda@gmail.com} \\
\And
  Shunbao Li \\
  School of Computer Science\\
  University of Sheffield\\
  Sheffield S10 2TN, UK \\
  \texttt{leeshunbao@gmail.com} \\
\And
  Zhipeng Yuan \\
  School of Computer Science\\
  University of Sheffield\\
  Western Bank, Sheffield S10 2TN, UK \\
  \texttt{zhipeng.yuan@sheffield.ac.uk} \\
\And
  Po Yang \\
  School of Computer Science\\
  University of Sheffield\\
  Sheffield S10 2TN, UK \\
  \texttt{po.yang@sheffield.ac.uk} \\
}

\begin{document}

\maketitle

\begin{abstract}
  Effective pest management is complex due to the need for accurate, context-specific decisions. Recent advancements in large language models (LLMs) open new possibilities for addressing these challenges by providing sophisticated, adaptive knowledge acquisition and reasoning. However, existing LLM-based pest management approaches often rely on a single-agent paradigm, which can limit their capacity to incorporate diverse external information, engage in systematic validation, and address complex, threshold-driven decisions. To overcome these limitations, we introduce PestMA, an LLM-based multi-agent system (MAS) designed to generate reliable and evidence-based pest management advice. Building on an editorial paradigm, PestMA features three specialized agents—an Editor for synthesizing pest management recommendations, a Retriever for gathering relevant external data, and a Validator for ensuring correctness. Evaluations on real-world pest scenarios demonstrate that PestMA achieves an initial accuracy of 86.8\% for pest management decisions, which increases to 92.6\% after validation. These results underscore the value of collaborative agent-based workflows in refining and validating decisions, highlighting the potential of LLM-based multi-agent systems to automate and enhance pest management processes. 
\end{abstract}

\section{Introduction}

Effective pest management is vital for maintaining agricultural productivity and safeguarding food security worldwide. Pests significantly reduce crop yields, pose health and environmental risks, and can lead to economic losses if not addressed promptly. Moreover, the escalating impacts of climate change, coupled with growing global demands for sustainable agriculture, call for adaptable and evidence-based strategies to prevent and mitigate pest infestations. Although various pest management approaches exist—including chemical controls, biological agents, and integrated pest management (IPM) programs—they often rely on specialized domain knowledge that is difficult to disseminate broadly or update regularly.

Recent advancements in large language models (LLMs) offer a novel means to consolidate and apply domain expertise more effectively. By leveraging these models, stakeholders can access comprehensive, up-to-date recommendations for informed pest management practices. In this paper, we introduce PestMA, a large language model (LLM)-based multi-agent system for generating reliable pest management advice. Our design draws on an editorial paradigm in which three agents—an Editor for synthesizing pest management advice (PMA), a Retriever for gathering relevant external knowledge to address the knowledge gaps in the Editor, and a Validator for checking PMA correctness—collaborate to address pest scenarios accurately. Evaluations on real-world pest data show that PestMA achieves 86.8\% accuracy for the synthesised pest management decision (PMD) and improves to 92.6\% following the validation from the Validator, demonstrating both the practical applicability of our approach and the enhancements gained from agent-based collaboration.

Our contributions are threefold:
\begin{itemize}
    \item Applying a Multi-Agent System to Pest Management: We propose PestMA, an LLM-based MAS explicitly tailored to address real-world pest management scenarios.
    \item Collaborative Agent Design for Pest Management: Our system features a unique Editor–Retriever–Validator workflow, illustrating how role specialization and collaboration enhance accuracy and reliability.
    \item Comprehensive PMD-Based Evaluation: We adopt the accuracy of the pest management decision (PMD) as a central metric to assess PestMA’s workflow and agent functionality, demonstrating that rigorous integration of external knowledge and logical validation significantly bolsters performance.
\end{itemize}

\section{Literature Review}

\subsection{LLM-based Multi-Agent Systems}

Recently, Large Language Models (LLMs)—such as GPT-4 \citep{achiam2023gpt}, Gemini-1.5 \citep{team2024gemini}, Llama-3 \citep{grattafiori2024llama}, Qwen-2.5 \citep{yang2024qwen2}, DeepSeek-R1 \citep{guo2025deepseek}—pretrained on substantial amounts of text have emerged as a universal foundation for addressing a broad range of AI tasks. To adapt these general-purpose LLMs to domain-specific contexts, Parameter-Efficient-Fine-Tuning (PEFT) techniques (e.g., LoRA \citep{hu2022lora}), along with prompt engineering \citep{sahoo2024systematic} and retrieval-augmented generation (RAG) \citep{lewis2020retrieval}, have achieved significant success. Driven by the remarkable capabilities demonstrated by LLMs, researchers have begun exploring their potential for undertaking more complex real-world applications, such as automated online purchases and research proposal generation.

This need has spurred the development of Agentic AI, referring to autonomous systems capable of accomplishing intricate tasks with minimal human intervention~\citep{bousetouane2025agentic, acharya2025agentic}. Owing to their strong language understanding and generative abilities, LLMs are increasingly viewed as pivotal performers of agentic AI. In particular, LLM-based multi-agent systems (MAS) have attracted attention for their capacity to exceed the capabilities of single-agent approaches~\citep{guo2024large, tran2025multi}. By enabling role specialization, supporting flexible interaction paradigms (e.g., collaboration or debate), and integrating diverse tools and skills, MAS can tackle more complex tasks than individual agents. This approach has proven effective in innovative applications such as Deep Research~\citep{openai2025deepresearch} by OpenAI—designed to generate entire research proposals—and Google’s AI co-scientist~\citep{gottweis2025ai}, which assists researchers by proposing novel scientific hypotheses. Hence, LLM-based MAS stands at the forefront of advanced agentic AI, demonstrating robust potential for tackling complex real-world tasks.

\subsection{LLM-based Pest Management}
Despite the growing success of LLMs in various fields, efforts to leverage these models for informed pest management have been relatively limited \citep{shaikh2024role}. Notably, \cite{yang2024gpt} employs GPT-based models to produce pest management advice. Conversely, \cite{yuan2024pestgpt} mitigates inherent knowledge gaps by infusing domain-specific information from reliable sources when prompting LLMs. However, none of the existing works extend to an LLM-based multi-agent system (MAS) for pest management. It is precisely this gap that the present research aims to address, by introducing a multi-agent framework where LLMs collaborate to provide comprehensive, validated pest management solutions.


\section{Methodology}
\label{sec: method}

We introduce PestMA (Pest Multi-Agents), an LLM-based multi-agent system structured around an editorial workflow, inspired by the processes typically found in news publication agencies. PestMA incorporates three distinct agents analogous to journalism roles:

\begin{itemize}
\item \textbf{Retriever (Journalist)}: Responsible for gathering relevant external data and documentation related to pest management.
\item \textbf{Editor}: Synthesizes the information provided by the Retriever into coherent, practical pest management advice (PMA).
\item \textbf{Validator (Managing Editor)}: Reviews and validates the synthesized advice to ensure credibility, reliability, and compliance with established standards and policies.
\end{itemize}

\begin{figure*}[th]
  \begin{center}
    \includegraphics[width=0.9\linewidth]{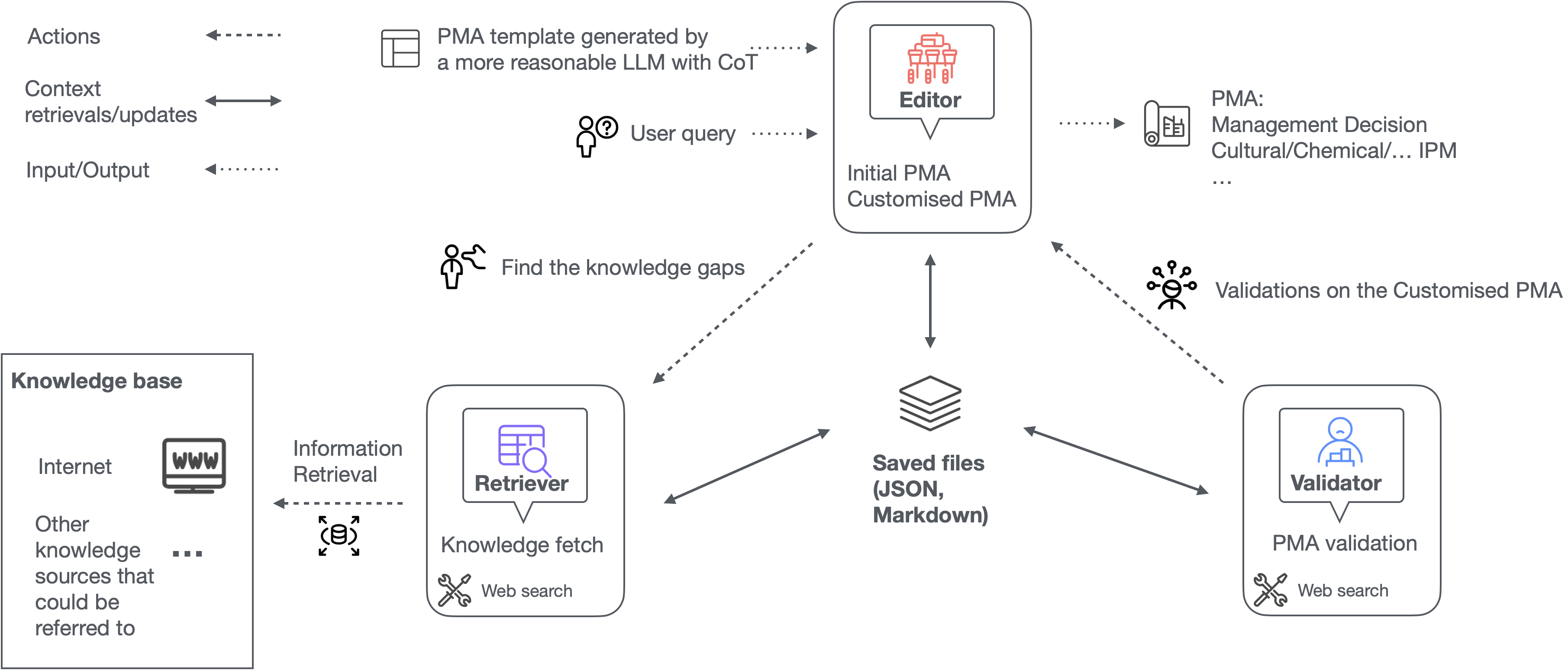}
  \end{center}
  \caption{The workflow of PestMA. Three agents--Editor, Retriever, and Validator--collaboratively generate the pest management advice given a pest management scenario requested by the user. Editor is responsible for synthesising the pest management advice, Retriever is tasked to find knowledge gaps in the PMA and search relevant information to fill the knowledge gaps, and the Validator accesses the PMA to ensure its accuracy and reliability.}
  \label{fig: workflow of pestma}
\end{figure*}

This structured editorial workflow ensures the integration of credible external knowledge and thorough validation processes, resulting in accurate, reliable, and actionable pest management recommendations. Figure~\ref{fig: workflow of pestma} illustrates the workflow of PestMA. The three agents mainly conduct the following tasks:

\paragraph{Initialising PMA.} Given an identified pest scenario, the Editor is tasked with generating an initial pest management advice (PMA). To ensure logical coherence and consistency, the Editor receives an initial prompt containing a PMA template generated by a reasoning model trained with Chain-of-Thought~\citep{wei2022chain} methods (e.g., GPT-o1~\citep{openai2024o1}). This template serves as structured guidance, facilitating logical analysis and eliciting the intrinsic knowledge of the Editor to produce a coherent initial PMA.

\paragraph{Customising PMA.} Once the initial PMA is generated, the Retriever identifies knowledge gaps that cannot be sufficiently addressed by the Editor’s intrinsic knowledge, such as local pesticide resistance levels. To address these gaps, the Retriever searches external sources, including relevant documentation and credible websites, to gather supplementary information tailored specifically to the pest scenario. The Retriever then summarizes these findings and returns them to the Editor, who subsequently integrates this external knowledge with the initial PMA. This integration results in a customised PMA that combines pretrained knowledge from the LLM with precise external insights, providing tailored advice specific to the user's scenario.

\paragraph{Validating PMA.} The Validator assesses the customised PMA to ensure its accuracy, credibility, and reliability. The Validator critically reviews the customised PMA produced by the Editor, identifying potential errors or inconsistencies, such as incorrect pest management decisions. Unlike the Retriever, who seeks external information to address knowledge gaps, the Validator utilizes external sources primarily for cross-validation and verification purposes. Thus, the Validator ensures robust validation of the customised PMA, enhancing the overall quality and reliability of the generated pest management advice.

\section{Experiments and Results}
\subsection{Dataset and Evaluation}
We evaluate the proposed PestMA on a dataset comprising 68 distinct pest management scenarios, covering 39 different pest species prevalent in the United Kingdom. Each scenario in the dataset consists of attributes of an identified pest scenario. In addition, a pest management decision (PMD) indicating if this scenario needs immediate pest management actions is attached to every scenario. The attached PMD is calculated by an expert system of pest management. Table \ref{tab: pest management scenario} depicts one sample of this dataset.

\begin{table}
  \caption{Attributes of an exemplary pest management scenario used in the PestMA evaluation}
  \label{tab: pest management scenario}
  \centering
  \begin{tabular}{ll}
    \toprule
    Attribute     & Value     \\
    \midrule
    Pest & Beet Cyst Nematode      \\
    Infestation Severity     & 1 egg and larvae per gram of soil     \\
    Crop Name     & Sugar Beet        \\
    Crop Growth Stage     & Seedling        \\
    Temperature & $15^\circ\text{C}$ \\ 
    Weather & Overcast \\ 
    Humidity & 75\% \\
    Precipitation & 20mm \\
    Time & April \\ 
    Location & Lincolnshire \\
    Pest Management Decision &  False \\

    \bottomrule
  \end{tabular}
\end{table}

As described in Section~\ref{sec: method}, we additionally introduce a PMA template to prompt PestMA. The PMA template used in our experiments is generated by the GPT-o1 model, a reasoning-oriented model trained using the CoT prompt combined with reinforcement learning. Guided by this PMA template, the initial PMA provided by the Editor agent addresses multiple aspects essential to effective pest management, including the pest management decision (PMD), integrated pest management (IPM) strategies, economic considerations, application timing, and post-treatment monitoring.

To evaluate PestMA systematically and efficiently, we adopt an ablation study approach by instructing the PestMA, specifically the Retriever and Validator, to focus on a single aspect in the initial PMA for customisation and validation. Focusing on one aspect prevents overwhelming the agents with excessive contextual complexity, thus reducing potential errors or hallucinations in agent responses. Additionally, concentrating on a single aspect significantly decreases the costs associated with extensive external information retrieval, such as frequent API calls on online search services provided by third parties.


Notably, we choose the pest management decision (PMD) as the focused aspect. PMD specifically involves a binary decision (true or false) indicating whether immediate action is necessary in response to the identified pest scenario requested by the user. Table~\ref{tab: pmd} describes how PMD is determined within PestMA. The decision is based on comparing infestation severity to an established threshold value. If the infestation severity exceeds this threshold, immediate management actions are advised, resulting in a true PMD; otherwise, the PMD is false. Accurate PMD decisions are crucial because inappropriate actions could negatively impact pest management effectiveness and cost-efficiency, thereby affecting PestMA’s credibility with users.


Evaluating PestMA based on the PMD allows us to assess both the collaboration among agents and the individual performance of each agent. Importantly, the threshold value is intentionally omitted from the initial prompts. Consequently, PestMA must independently retrieve this threshold from reliable external knowledge sources, such as data provided by the Agriculture and Horticulture Development Board (AHDB) or the British Crop Production Council (BCPC). The Editor leverages intrinsic knowledge to initially estimate PMD, while the Retriever locates and retrieves the exact threshold from external databases. Finally, the Validator ensures the correctness of the comparison made by the Editor, thus verifying the reliability of the final PMD.


Given that PMD is a binary classification task, we utilize accuracy as our evaluation metric. Accuracy is calculated using the following formula:

\begin{equation}
    \text{Accuracy} = \frac{\text{Number of Correct Predictions}}{\text{Total Number of Predictions}}
    \label{eq: accuracy metrics}
\end{equation}

\begin{table}
  \caption{Illustration of determining Pest Management Decision (PMD) based on infestation severity relative to an established threshold}
  \label{tab: pmd}
  \centering
  \begin{tabular}{ll}
    \toprule
    Description     & Value     \\
    \midrule
    Pest & Beet Cyst Nematode      \\
    Crop Name     & Sugar Beet        \\
    Crop Growth Stage     & Seedling        \\
    Infestation Severity     & 1 egg and larvae per gram of soil     \\
    Threshold & 2 eggs and larvae per relevant soil volume (source: AHDB) \\ \midrule
    PMD & False \\

    \bottomrule
  \end{tabular}
\end{table}

\subsection{Agent and Task Profiling}

We developed PestMA using the CrewAI framework, which enables explicit agent and task profiling. Table \ref{tab: agent profiling} presents an overview of each agent’s profile and the tasks assigned to them. Tables \ref{tab: task profiling 1} and \ref{tab: task profiling 2} outline the specific tasks within PestMA and detail the tools that support these tasks. The profiling of the agents and tasks define the collaborative workflow for PestMA described in Section~\ref{sec: method}, with a particular focus on the PMD aspect in the PMA.

\begin{table}
\caption{Agent profiles and their assigned tasks in PestMA}
\label{tab: agent profiling}
\centering
\begin{tabular}{
    >{\centering}m{1.5cm}  
    >{}m{6cm}    
    >{\centering\arraybackslash}m{5cm}    
}
\toprule
Agent & \centering Profile & Related Tasks \\
\midrule
Editor & You are a agronomist with exception knowledge about pest management. Coordinate and synthesize information to generate both initial and final pest management advice (PMA) to the queried pest identification scenario.  & 
\begin{itemize}
  \item Generate initial PMA.
  \item Generate customised PMA. 
\end{itemize} \\
\addlinespace
Retriver & Review the initial PMA created by the Editor and suggest a customisation plan to enhance the initial PMA. With the customisation plan, retrieve detailed online information, using the provided search queries and recommended online sources. & 
\begin{itemize}
  \item Make customisation plan. 
  \item Knowledge retrieval. 
\end{itemize} \\
\addlinespace
Validator & Critically review and validate the customised PMA, with a specific emphasis on verifying that the threshold conclusion is accurately derived from the diagnostic data. Leverage both your intrinsic expertise and external online search tool to ensure that the decision about whether the pest infestation exceeds the action threshold is scientifically sound and consistent with industry guidelines. & 
\begin{itemize}
  \item Validate threshold. 
\end{itemize} \\
\bottomrule
\end{tabular}
\end{table}

\begin{table}
\caption{Task definitions and the tools used by each agent (Part 1)}
\label{tab: task profiling 1}
\centering
\begin{tabular}{
    >{\centering}m{2cm}  
    >{}m{6cm}    
    >{\centering\arraybackslash}m{3cm}    
}
\toprule
Agent & \centering Profile & Related Tools \\
\midrule
Generate initial PMA & Generate comprehensive pest management advice (PMA) on a pest identification scenario queried by the user. The request scenario is detailed in a JSON file at \{query\_path\}.
    To generate the PMA to the queried scenario, an exemplary scenario and a PMA template to this exemplary scenario are provided for references.
    Particularly, the exemplary scenario is detailed in the same format to the query scenario and can be read at \{example\_path\}. Read it first.
    The PMA template to the exemplary scenario is a markdown file and can be read at \{example\_pma\_path\}. Read this PMA after reading the exemplary scenario.
    The generated PMA to the queried scenario should follow the pattern of the exemplary PMA template.
    The task involves reading JSON files with JSON reader and the markdown file using MDReader, parsing their sections (pest identification, control methods, preventative measures, and environmental considerations), and populating them with detailed, queried PMA.  & 
\begin{itemize}
  \item JSON reader.
  \item Markdown reader.
\end{itemize} \\
\addlinespace
    Make customisation plan & Review the initial PMA saved as a Markdown file at \{initial\_pma\_path\} and identify the gaps in the pest management decision, which could be filled by searching online knowledge,
    therefore leading to a customised PMA based on the queried scenario.
    Compile the identified gaps into sections. For each section, list the necessary search queries and recommended online sources,
    along with a justification for the recommendation.
    For the online sources, they must be trustworthy and relevant to the region queried.
    For instances, sources from Agriculture and Horticulture Development Board (AHDB), EU-FarmBook, and British Crop Production Council (BCPC). & 
\begin{itemize}
  \item Markdown reader.
\end{itemize} \\
\bottomrule
\end{tabular}
\end{table}

\begin{table}
\caption{Task definitions and the tools used by each agent (Part 2)}
\label{tab: task profiling 2}
\centering
\begin{tabular}{
    >{\centering}m{2cm}  
    >{}m{6cm}    
    >{\centering\arraybackslash}m{3cm}    
}
\toprule
Agent & \centering Profile & Related Tools \\
\midrule
Knowledge retrieval & 
    Retrieve detailed online information for each enhancement section specified in the customization plan.
    The plan is a JSON file that can be read at \{custom\_plan\_path\}.
    For each section in the plan, use the provided search queries and recommended online sources
    to gather relevant information. Compile a comprehensive report that analyses and summary the information searched. The report should be composed with enhancement sections in accordance with the customization plan. Each enhancement section should include: 1. The section name. 2. A summary of findings for each search query. 3. Relevant citations\&links from the recommended online sources. 4. An analysis that highlights how the new data can enhance the initial PMA. & 
\begin{itemize}
  \item JSON reader.
  \item Online search.
\end{itemize} \\
\addlinespace
    Generate customised PMA &  
    Using the initial PMA generated in \{initial\_pma\_path\} and the report summarised from online information by the Retriever in \{retrieved\_info\_path\},
    synthesize a final, customised pest management advice (PMA) for the queried scenario.
    This task involves integrating the additional insights (addressing the sections identified in the customization plan in \{custom\_plan\_path\}) into the initial PMA. Ensure that the final document adheres to the structure and style of the initial PMA.& 
\begin{itemize}
    \item JSON reader.
  \item Markdown reader.
\end{itemize} \\
\addlinespace
    Validate threshold &  
    Analyze the customised PMA (located at \{custom\_pma\_path\}) with a focus on the threshold conclusion and management (or action) decision.
    Examine whether the diagnostic data—such as the pest identification details and infestation levels—logically supports the conclusion stated under "Threshold Exceeded."
    Identify any discrepancies or areas that may require further external validation. In cases where the threshold evaluation reveals uncertainty or potential misalignment with standard practices,
    initiate searches for updated industry guidelines, scientific literature, or official thresholds related to pest infestation levels. Compare the external findings with the customised PMA’s threshold decision to validate or recommend corrections. & 
\begin{itemize}
  \item JSON reader.
  \item Markdown reader.
  \item Online search.
\end{itemize} \\
\bottomrule
\end{tabular}
\end{table}

\subsection{Results}
\label{sec: results}

We evaluated the accuracy of PestMA’s pest management decision (PMD) at two distinct stages. First, we measured the accuracy of the PMD proposed by the Editor after integrating information retrieved by the Retriever. This stage reflects the collaboration and effectiveness of both the Editor and Retriever agents. Second, we assessed the PMD validated by the Validator, indicating whether the Validator can accurately confirm (or correct) the logic used by the Editor when comparing infestation severity to the threshold.

Table \ref{tab: test performance} reports the accuracy for both stages. Notably, the Editor’s PMD in the customised PMA achieves an accuracy of 86.8\%, which is already a strong baseline performance. Once validated, the PMD accuracy further increases to 92.6\%, demonstrating that the Validator can detect and correct threshold comparison errors, thereby enhancing PestMA’s overall reliability.


Table \ref{tab: a test example} presents a representative test example illustrating how the Validator safeguards the PMD. In this scenario, the Validator identifies that the Editor’s decision, while reasonably justified, does not explicitly reference the threshold established by regional guidelines. By incorporating guidance from the AHDB, the Validator clarifies that immediate action is still recommended when infestation closely approaching to the threshold of a nematode count of 1,000 per litre of soil. This process highlights the Validator’s role in reinforcing evidence-based recommendations, ultimately resulting in improved decision accuracy and greater confidence in the response of PestMA.

\begin{table}
\caption{PMD accuracy at different stages of PestMA’s workflow}
\label{tab: test performance}
\centering
\begin{tabular}{@{}ll@{}}
\toprule
Response stage in PestMA & Acc. (\%)  \\ \midrule
PMD (Editor + Retriever) &  86.8  \\ 
PMD (Validated by Validator) &  92.6 \\

\bottomrule
\end{tabular}
\label{tab: PMD accuracy}
\end{table}

\begin{table}
\caption{Example scenario demonstrating how the Validator refines the PMD}
\label{tab: a test example}
\centering
\begin{tabular}{
    >{}m{3cm}  
    >{}m{3cm}    
    >{}m{6cm}    
}
\toprule
 Scenario & PMD in customised PMA & Validated PMD \\
\midrule
    \textit{Pest}: Free-Living Nematodes,
    \textit{InfestationSeverity}: \textcolor{red}{800} Trichodorus nematodes per litre of soil,
    \textit{CropName}: Sugar Beet,
    \textit{CropGrowthStage}: Early root development,
    \textit{Temperature}: 12°C,
    \textit{Weather}: Partly cloudy,
    \textit{Humidity}: 75\%,
    \textit{Precipitation}: 30 mm,
    \textit{Time}: May,
    \textit{Location}: Norfolk, 
    & With \textcolor{red}{800} nematodes per litre of soil, consider intervention due to potential root damage and yield loss during this vital growth phase.  
    & AHDB Guidance: According to the AHDB, severe symptoms in sugar beet can occur when Trichodorus populations exceed \textcolor{red}{1,000} nematodes per litre of soil. Hence, \textcolor{red}{interventions often are recommended when levels approach this threshold} (source: [AHDB Knowledge Library](https://ahdb.org.uk/knowledge-library/free-living-nematodes-and-their-impact-on-the-yield-and-quality-of-field-crops)).
    The inital PMD asserts an infestation of 800 nematodes per litre of soil, aligning closely with regional guidelines suggesting severe symptoms develop above 1,000 nematodes per litre. \textcolor{red}{While the PMD emphasizes timely intervention, it would strengthen its recommendations by explicitly correlating with the 1,000 nematode threshold mentioned in regional studies.} \\
\bottomrule
\end{tabular}
\end{table}

\section{Conclusion}

In this paper, we introduced PestMA, a multi-agent system (MAS) underpinned by large language models (LLMs) for delivering precise and reliable pest management advice. By structuring the MAS with an editorial paradigm—an Editor synthesizing pest management advice, a Retriever gathering external domain knowledge, and a Validator verifying correctness—we demonstrated how specialized roles and agent-based collaboration can enhance decision accuracy. Our empirical results on real-world pest data showed that PestMA’s customised pest management decision (PMD) attains 86.8\% accuracy, which is further improved to 92.6\% following the Validator’s intervention.

Despite these promising outcomes, two limitations remain. First, PestMA currently relies exclusively on an online search tool for external knowledge retrieval; future work can integrate more advanced approaches, such as retrieval-augmented generation (RAG) and other domain-specific tools. Second, our evaluation focused specifically on the PMD dimension, thus providing a narrower scope of assessment. In forthcoming research, PestMA will be extended to encompass additional aspects of pest management, ensuring a more holistic evaluation and further strengthening the system’s practical utility.

\bibliographystyle{plainnat}
\bibliography{references}

\begin{thebibliography}{19}
\providecommand{\natexlab}[1]{#1}
\providecommand{\url}[1]{\texttt{#1}}
\expandafter\ifx\csname urlstyle\endcsname\relax
  \providecommand{\doi}[1]{doi: #1}\else
  \providecommand{\doi}{doi: \begingroup \urlstyle{rm}\Url}\fi

\bibitem[Acharya et~al.(2025)Acharya, Kuppan, and Divya]{acharya2025agentic}
Deepak~Bhaskar Acharya, Karthigeyan Kuppan, and B~Divya.
\newblock Agentic ai: Autonomous intelligence for complex goals--a comprehensive survey.
\newblock \emph{IEEE Access}, 2025.

\bibitem[Achiam et~al.(2023)Achiam, Adler, Agarwal, Ahmad, Akkaya, Aleman, Almeida, Altenschmidt, Altman, Anadkat, et~al.]{achiam2023gpt}
Josh Achiam, Steven Adler, Sandhini Agarwal, Lama Ahmad, Ilge Akkaya, Florencia~Leoni Aleman, Diogo Almeida, Janko Altenschmidt, Sam Altman, Shyamal Anadkat, et~al.
\newblock Gpt-4 technical report.
\newblock \emph{arXiv preprint arXiv:2303.08774}, 2023.

\bibitem[Bousetouane(2025)]{bousetouane2025agentic}
Fouad Bousetouane.
\newblock Agentic systems: A guide to transforming industries with vertical ai agents.
\newblock \emph{arXiv preprint arXiv:2501.00881}, 2025.

\bibitem[Gottweis and Natarajan(2025)]{gottweis2025ai}
Juraj Gottweis and Vivek Natarajan.
\newblock Accelerating scientific breakthroughs with an ai co-scientist.
\newblock \url{https://research.google/blog/accelerating-scientific-breakthroughs-with-an-ai-co-scientist/}, February 2025.
\newblock Accessed: 2025-04-11.

\bibitem[Grattafiori et~al.(2024)Grattafiori, Dubey, Jauhri, Pandey, Kadian, Al-Dahle, Letman, Mathur, Schelten, Vaughan, et~al.]{grattafiori2024llama}
Aaron Grattafiori, Abhimanyu Dubey, Abhinav Jauhri, Abhinav Pandey, Abhishek Kadian, Ahmad Al-Dahle, Aiesha Letman, Akhil Mathur, Alan Schelten, Alex Vaughan, et~al.
\newblock The llama 3 herd of models.
\newblock \emph{arXiv preprint arXiv:2407.21783}, 2024.

\bibitem[Guo et~al.(2025)Guo, Yang, Zhang, Song, Zhang, Xu, Zhu, Ma, Wang, Bi, et~al.]{guo2025deepseek}
Daya Guo, Dejian Yang, Haowei Zhang, Junxiao Song, Ruoyu Zhang, Runxin Xu, Qihao Zhu, Shirong Ma, Peiyi Wang, Xiao Bi, et~al.
\newblock Deepseek-r1: Incentivizing reasoning capability in llms via reinforcement learning.
\newblock \emph{arXiv preprint arXiv:2501.12948}, 2025.

\bibitem[Guo et~al.(2024)Guo, Chen, Wang, Chang, Pei, Chawla, Wiest, and Zhang]{guo2024large}
Taicheng Guo, Xiuying Chen, Yaqi Wang, Ruidi Chang, Shichao Pei, Nitesh~V Chawla, Olaf Wiest, and Xiangliang Zhang.
\newblock Large language model based multi-agents: A survey of progress and challenges.
\newblock \emph{arXiv preprint arXiv:2402.01680}, 2024.

\bibitem[Hu et~al.(2022)Hu, Shen, Wallis, Allen-Zhu, Li, Wang, Wang, Chen, et~al.]{hu2022lora}
Edward~J Hu, Yelong Shen, Phillip Wallis, Zeyuan Allen-Zhu, Yuanzhi Li, Shean Wang, Lu~Wang, Weizhu Chen, et~al.
\newblock Lora: Low-rank adaptation of large language models.
\newblock \emph{ICLR}, 1\penalty0 (2):\penalty0 3, 2022.

\bibitem[Lewis et~al.(2020)Lewis, Perez, Piktus, Petroni, Karpukhin, Goyal, K{\"u}ttler, Lewis, Yih, Rockt{\"a}schel, et~al.]{lewis2020retrieval}
Patrick Lewis, Ethan Perez, Aleksandra Piktus, Fabio Petroni, Vladimir Karpukhin, Naman Goyal, Heinrich K{\"u}ttler, Mike Lewis, Wen-tau Yih, Tim Rockt{\"a}schel, et~al.
\newblock Retrieval-augmented generation for knowledge-intensive nlp tasks.
\newblock \emph{Advances in neural information processing systems}, 33:\penalty0 9459--9474, 2020.

\bibitem[OpenAI(2024)]{openai2024o1}
OpenAI.
\newblock Introducing openai o1.
\newblock \url{https://openai.com/o1/}, 2024.
\newblock Accessed: 2025-04-10.

\bibitem[{OpenAI}(2025)]{openai2025deepresearch}
{OpenAI}.
\newblock Introducing deep research.
\newblock \url{https://openai.com/index/introducing-deep-research/}, February 2025.
\newblock Accessed: 2025-04-11.

\bibitem[Sahoo et~al.(2024)Sahoo, Singh, Saha, Jain, Mondal, and Chadha]{sahoo2024systematic}
Pranab Sahoo, Ayush~Kumar Singh, Sriparna Saha, Vinija Jain, Samrat Mondal, and Aman Chadha.
\newblock A systematic survey of prompt engineering in large language models: Techniques and applications.
\newblock \emph{arXiv preprint arXiv:2402.07927}, 2024.

\bibitem[Shaikh et~al.(2024)Shaikh, Rasool, Veningston, and Yaseen]{shaikh2024role}
Tawseef~Ayoub Shaikh, Tabasum Rasool, K~Veningston, and Syed~Mufassir Yaseen.
\newblock The role of large language models in agriculture: harvesting the future with llm intelligence.
\newblock \emph{Progress in Artificial Intelligence}, pages 1--48, 2024.

\bibitem[Team et~al.(2024)Team, Georgiev, Lei, Burnell, Bai, Gulati, Tanzer, Vincent, Pan, Wang, et~al.]{team2024gemini}
Gemini Team, Petko Georgiev, Ving~Ian Lei, Ryan Burnell, Libin Bai, Anmol Gulati, Garrett Tanzer, Damien Vincent, Zhufeng Pan, Shibo Wang, et~al.
\newblock Gemini 1.5: Unlocking multimodal understanding across millions of tokens of context.
\newblock \emph{arXiv preprint arXiv:2403.05530}, 2024.

\bibitem[Tran et~al.(2025)Tran, Dao, Nguyen, Pham, O'Sullivan, and Nguyen]{tran2025multi}
Khanh-Tung Tran, Dung Dao, Minh-Duong Nguyen, Quoc-Viet Pham, Barry O'Sullivan, and Hoang~D Nguyen.
\newblock Multi-agent collaboration mechanisms: A survey of llms.
\newblock \emph{arXiv preprint arXiv:2501.06322}, 2025.

\bibitem[Wei et~al.(2022)Wei, Wang, Schuurmans, Bosma, Xia, Chi, Le, Zhou, et~al.]{wei2022chain}
Jason Wei, Xuezhi Wang, Dale Schuurmans, Maarten Bosma, Fei Xia, Ed~Chi, Quoc~V Le, Denny Zhou, et~al.
\newblock Chain-of-thought prompting elicits reasoning in large language models.
\newblock \emph{Advances in neural information processing systems}, 35:\penalty0 24824--24837, 2022.

\bibitem[Yang et~al.(2024{\natexlab{a}})Yang, Yang, Zhang, Hui, Zheng, Yu, Li, Liu, Huang, Wei, et~al.]{yang2024qwen2}
An~Yang, Baosong Yang, Beichen Zhang, Binyuan Hui, Bo~Zheng, Bowen Yu, Chengyuan Li, Dayiheng Liu, Fei Huang, Haoran Wei, et~al.
\newblock Qwen2. 5 technical report.
\newblock \emph{arXiv preprint arXiv:2412.15115}, 2024{\natexlab{a}}.

\bibitem[Yang et~al.(2024{\natexlab{b}})Yang, Yuan, Li, Peng, Liu, and Yang]{yang2024gpt}
Shanglong Yang, Zhipeng Yuan, Shunbao Li, Ruoling Peng, Kang Liu, and Po~Yang.
\newblock Gpt-4 as evaluator: Evaluating large language models on pest management in agriculture.
\newblock \emph{arXiv preprint arXiv:2403.11858}, 2024{\natexlab{b}}.

\bibitem[Yuan et~al.(2024)Yuan, Liu, Peng, Li, Leybourne, Musa, Huang, and Yang]{yuan2024pestgpt}
Zhipeng Yuan, Kang Liu, Ruoling Peng, Shunbao Li, Daniel Leybourne, Nasamu Musa, He~Huang, and Po~Yang.
\newblock Pestgpt: Leveraging large language models and iot for timely and customized recommendation generation in sustainable pest management.
\newblock \emph{IEEE Internet of Things Magazine}, 8\penalty0 (1):\penalty0 26--33, 2024.

\end{thebibliography}

\end{document}